\documentclass[runningheads]{llncs}
\usepackage{graphicx}
\usepackage{xcolor}
\usepackage{comment}
\usepackage{subcaption}
\usepackage{graphicx}
\usepackage{multirow}
\usepackage{floatrow}
\newfloatcommand{capbtabbox}{table}[][\FBwidth]
\usepackage{blindtext}

\newcommand{\tool}[0]{\mbox{\textsc{ComScribe}}} 

\DeclareFontFamily{\encodingdefault}{\ttdefault}{\hyphenchar\font=`\-} 
\begin{document}
\title{Monitoring Collective Communication\\ Among GPUs}

%
%
\author{Muhammet Abdullah Soytürk \orcidID{0000-0002-2880-0857} \and 
\\Palwisha Akhtar
\orcidID{0000-0003-0279-031X} \and
Erhan Tezcan
\orcidID{0000-0001-5129-4166}
\and 
Didem Unat
\orcidID{0000-0002-2351-0770}
}

%
\authorrunning{M. Soyturk et al.}
%
\institute{Department of Computer Science and Engineering, Ko\c{c}
 University, Turkey \\ 
\email{\{msoyturk20,pakhtar19,etezcan19,dunat\}@ku.edu.tr}}
\maketitle
\vspace{-0.2cm}
\begin{abstract}
Communication among devices in multi-GPU systems plays an important role in terms of performance and scalability. In order to optimize an application,  programmers need to know the type and amount of the communication happening among GPUs. Although there are prior works to gather this information in MPI applications on distributed systems and multi-threaded applications on shared memory systems,
there is no tool that identifies communication among GPUs. 
Our prior work, \tool{}, presents a point-to-point (P2P) communication detection tool for GPUs sharing a common host. In this work, we extend \tool{} to identify communication among GPUs for collective and P2P communication primitives in NVIDIA's NCCL library. In addition to P2P communications, collective communications are commonly used in HPC and AI workloads thus it is important to monitor the induced data movement due to collectives. Our tool extracts the size and the frequency of data transfers in an application and visualizes them as a communication matrix. To demonstrate the tool in action, we present communication matrices and some statistics for two applications coming from machine translation and image classification domains. 

\end{abstract}
\vspace{-0.5cm}

\section{Introduction}
\label{sec: Intro}
Nowadays, multi-GPU systems are commonly employed for parallel applications either to reduce execution time or to enable processing a large amount of data. In a multi-GPU application, there are many alternative ways for devices to communicate, thus choosing the right communication type can become a critical performance contributor. In convolutional neural networks (CNNs), for instance, while data and spatial parallelism based implementations may perform gradient exchange at the end of each iteration, filter and channel parallelism based implementations may require multiple collective communication calls at each layer \cite{oracle}, resulting different performance behaviour and scalability. Hence, identifying the type and size of the communication among GPUs can guide the programmer in many aspects for performance optimization.       

Broadly speaking, communication on a multi-processor system can be categorized into two types: P2P communication between two processors (e.g., GPUs) or collective communication among multiple processors. For P2P communication, CUDA API offers various data transfer schemes to the programmer by utilizing Unified Virtual Addressing (UVA), Zero-copy Memory and Unified Memory paradigms. For collective communication, NVIDIA offers NCCL \cite{nccl} library which provides efficient and topology-aware collectives. Collective primitives are used in various parallel algorithms that require collective work done by a group of processors. For example, many deep learning applications require data to be distributed in many processors and share the gradients among themselves, typically with an All-Reduce collective. Hence, deep learning frameworks such as PyTorch, Tensorflow and MxNet have already integrated NCCL into their frameworks to perform collective calls \cite{nccl}.

Communication monitoring among GPUs can help reason about scalability issues and performance divergence between different implementations of the same application, and guide the programmer to utilize the interconnects for better performance. For instance, if a single GPU application is scaled up to multiple GPUs, it may follow a master-slave communication pattern, which would underutilize the GPU interconnects.
Because of the aforementioned reasons, identifying the volume of communication for different communication patterns offer avenues to improve performance and tune software for scalability.

To the best of our knowledge, there is no communication monitoring tool for NCCL collective communication primitives in multi-GPU systems. Previous work on communication monitoring includes identification of MPI collectives on distributed systems such as EZTrace \cite{eztrace}. EZTrace can identify explicit P2P communication functions that CUDA offers such as cudaMemcpy but it cannot identify Unified Memory, Zero-Copy memory and NCCL collective communication primitives. Similarly, NVIDIA's profiler \textit{nvprof} \cite{nvprof} cannot provide any information about data transfers in NCCL primitives because data movement in NCCL is not based on \textit{cudaMemcpy} API. \textit{Nsight Systems} \cite{Nsight}, a system-wide performance analysis tool by NVIDIA, visualizes the timeline of collective calls together with other kernel information but does not present overall picture of the data movement. Moreover, it does not provide any visual or machine readable data on the amount of data movement between GPU pairs. 


This work  extends \tool{} \cite{ComScribe}, a tool that can monitor, identify, and quantify different types of communication among GPU devices, to support collective communication primitives. 
\tool{} can extract communication-related activities in an application and generate a communication matrix that shows the amount of data movement between GPU-GPU or GPU-CPU pairs. It leverages the NVIDIA's profiling tool \textit{nvprof} to monitor P2P communication. However, a significantly different approach is required to monitor collective communications because \textit{nvprof} is not capable of providing any information about NCCL collectives. Our extension to the \tool{} tool  overcomes this limitation and works in three steps: First, we preload the NCCL library with extra functionality for logging the data transfers. Second, we collect GPU-GPU memory transfer information during the execution. Finally, we perform post-processing to quantify communication among GPUs and generate the communication matrices. 
Our contributions are summarized below: 

\begin{itemize}
     \item[$-$] We extend \tool{} to provide a more complete coverage of the communication types and monitor data transfers between GPUs during the execution of collective communication primitives. 
    \item[$-$] We present communication statistics and communication matrices for a machine translation and an image classification applications to demonstrate how \tool{} can be used for explaining different implementations of data parallelism.
    \item The extensions are incorporated in \tool{}, which is available at \\ \url{https://github.com/ParCoreLab/ComScribe}.
\end{itemize}

The rest of the paper is organized as follows. In Section 2, we discuss the previous work on P2P communication monitoring with \tool{} and introduce NVIDIA Collective Communication Library (NCCL). It also explains all NCCL collective communication primitives. In Section 3, we discuss the design and implementation of collective communication monitoring. Section 4 shows the results on selected applications. Section 5 describes the related work. Section 6 presents our conclusions.
\vspace{-0.4cm}
\section{Background}
\label{background}
In this section, we first introduce the previous work on point-to-point communication monitoring with \tool{}. Then, we discuss the collective communication primitives supported by the NCCL.

\subsection{Point-to-Point Communication Monitoring with \tool{}}

\tool{} was originally developed to identify P2P communication of host-device and device-device pairs for various data transfer types offered by CUDA APIs. It supports the monitoring of explicit data transfers such as \texttt{cudaMemcpy} as well as implicit data transfers such as Zero-Copy Memory and Unified Memory. It is implemented on top of NVIDIA's profiling tool \textit{nvprof}, which can generate intra-node P2P communication information together with computation-related information in a machine readable format. Once the necessary profiling data is generated, \tool{} extracts the relevant information and generates communication matrices.

\subsubsection{Host-Device Communication.}
In CUDA programming, a memory transfer between a host and a device can be realized in two ways: explicit transfer and implicit transfer. An explicit transfer refers to the \texttt{cudaMemcpy} or \texttt{cudaMemcpyAsync} function in CUDA Runtime API where the programmer can explicitly specify the kind (Host-to-Device, Device-to-Host, or cudaMemcpyDefault) of the memory transfer. Implicit transfer types are Zero-Copy memory and Unified Memory. Zero-Copy memory paradigm allows a GPU to directly access host memory over PCIe or NVLink interconnect by pinning a memory region in host memory and mapping it to the GPU. A memory region allocated with Unified Memory via \texttt{cudaMallocManaged} is accessible from any processor (CPU or GPU) in the system. Page faults are handled by the page migration engine automatically. 
\vspace{-0.2cm}
\subsubsection{Device-Device Communication.}
As in host-device communication, there are two types of a data transfer: explicit transfers and implicit transfers. In an explicit transfer, the programmer can use either \texttt{cudaMemcpy} or \texttt{cudaMemcpyPeer}. If peer access is disabled, the data will be copied to the host and then transferred to the destination device. In P2P communication, implicit transfer types are also Zero-Copy memory or Unified Memory. In Zero-Copy memory, devices with peer access capability can read and write to each others’ memory through the data pointer. In Unified Memory, any memory region allocated with \texttt{cudaMallocManaged} can be accessed by the peer GPUs.   
\subsection{NCCL for GPU-based Collective Communication}
NCCL is NVIDIA's Collective Communications Library that provides efficient and topology-aware inter-GPU communication. It implements both collective and point-to-point communication primitives for intra-node and inter-node communication. NCCL has the ability to detect and utilize various interconnects such as PCIe, NVLINK, InfiniBand Verbs, and IP sockets. This feature eliminates the burden of optimizing applications for systems with different topology or interconnects. 

Collective communication involves a data transfer between more than one GPU, unlike P2P communication where there is only one sender and receiver. In order to use a collective primitive on a group of GPUs (i.e. in a communicator), each GPU within the communicator is assigned a zero-based rank and each rank involved in a collective communication must call the same communication primitive function with compatible arguments. For example, they must be in the same communicator. 

The need for efficient implementation of collective communication primitives comes from the fact that many parallel algorithms share data among a group of processors (i.e., communicator). Especially, the need for abundance of data in deep learning models require data to be distributed in many processors and share the gradients among processors, typically with an All-Reduce collective. Hence, deep learning frameworks such as PyTorch, Tensorflow and MxNet have already integrated NCCL into their frameworks to perform collective calls. 


Before the advent of NCCL, collective primitives would be implemented through a combination of CUDA memory copy operations and CUDA kernels for local reductions. In NCCL, each collective is implemented in a single kernel that handles both communication and computation operations in order to speed up the synchronization and minimize the resources needed to reach peak bandwidth.

\subsubsection{Collective Communication Primitives.}

NCCL provides five collective communication primitives: Broadcast, Reduce, ReduceScatter, AllGather, and AllReduce. Especially, AllReduce is frequently used in deep learning applications to share the local gradients among processors. NCCL's collective communication primitives are similar to MPI's collective communication primitives. The functionality of each collective primitive is described below:  
\begin{itemize}
\item Broadcast: The Broadcast collective copies data buffer that resides in the root rank's memory to the all other ranks. 

\item Reduce:
The Reduce collective performs a reduction operation on data (e.g. sum, max) aggregated from all ranks in a communicator and writes the result in the specified rank.

\item ReduceScatter:
The ReduceScatter collective performs the same operation as the Reduce operation, except the result is scattered in equal blocks among ranks, each rank getting a chunk of data based on its rank index. 

\item AllGather:
In AllGather, each rank in the communicator aggregates $N$ values from every rank into an output buffer. The output is ordered by rank index.

\item AllReduce: The AllReduce collective is similar to the Reduce collective. The only functional difference is that the result of the reduction is written into each rank's receive buffer in the communicator instead of one rank. AllReduce is a rank agnostic operation, i.e. reordering of ranks does not affect the outcome since all ranks will have identical data at the end. This operation is functionally equivalent to a Reduce followed by a Broadcast.
\end{itemize}

\subsubsection{Point-to-Point Primitives.} P2P primitives (ncclSend, ncclRecv) were added to NCCL 2.7. These primitives allow users to express primitives that are not directly implemented in NCCL such as one-to-all (scatter), all-to-one (gather), and all-to-all communication operations.


\vspace{-0.4cm}


\section{Collective Communication Monitoring}
\label{sec:design}


In \tool{}, design of collective communication monitoring is significantly different than P2P communication monitoring. \tool{} leverages \textit{nvprof} to capture P2P communication information to construct the communication matrices. However, this approach  is not applicable to collective communication monitoring because \textit{nvprof} does not provide any memory transfer information about NCCL collective primitives. NVIDIA's new profiling tool Nsight Systems could serve as an alternative approach for NCCL profiling but even though it can visualize the execution timeline of NCCL kernels, it does not provide any information on data transfers in a machine readable format. Moreover, the information provided by Nsight Systems is convoluted with the compute kernel information required for the collective primitives, which makes it hard for the programmer to distill the communication related activities. 

\begin{figure}[t]
\begin{center}
  \includegraphics[width=0.8\textwidth]{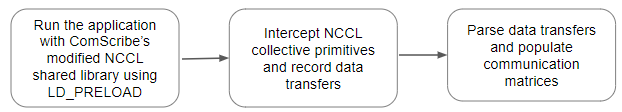}
  \end{center}
  \caption{Workflow diagram of \tool{}}
  \label{fig:workflow}
\vspace{-0.5cm}
\end{figure}

Figure \ref{fig:workflow} illustrates the collective communication monitoring workflow added to \tool{}. \tool{} employs LD\_PRELOAD utility to intercept NCCL calls and records the data transfers of collective primitives. The main benefit of this approach is that it eliminates the need to change the source code of the binary being investigated by the user.  

In order to use NCCL, the initialization step is to create a communicator and allocate a send buffer and a receive buffer for each device.
Creation of the communicator involves generating a unique id for the communicator and assigning zero-based rank to each device in the communicator. After the initialization, the programmer can make multiple collective calls on the communicator. The same collective call must be performed by each rank in the communicator. 

Internally, NCCL computes the data transfer channels and decides on which algorithm to be used based on the estimation of how long each algorithm would take for each collective call and enqueues the work to a queue. \tool{} retrieves this data before the execution of the collective call on the devices. At the end of the execution, \tool{} generates a single output file that contains the data transfers of each device in the communicator. Then, it parses these output files and generates communication matrices and other communication related statistics. 

\subsubsection{AllReduce.} While NCCL implements Broadcast, Reduce, AllGather and ReduceScatter operations with only ring algorithm, it provides three algorithms for AllReduce: ring, tree and collnet. The algorithm used for is important for profiling because it affects the amount of communication among ranks. Table \ref{tab:algotable} shows the data movement induced by each algorithm.

\setlength{\tabcolsep}{1em}
\begin{table}[h]
\vspace{-0.3cm}
\caption{\label{tab:algotable}Number of bytes sent and received by a rank in the communicator for AllReduce operation. $S$ is the size of the data, $N$ is the number of ranks}
\begin{center}
\begin{tabular}{ |c|c|c| } 
 \hline
 Algorithm Types & Intranode & Internode \\ 
 \hline
 Ring & $2 \times (N-1) \times S/N$ & $2 \times (N-1) \times S/N$ \\
 \hline
 Tree & root: $S$, others: $2 \times S$ & root: $S$, others: $2 \times S$ \\
  \hline
 Collnet & $2 \times S$ & $S$ \\ 
 \hline
\end{tabular}
\end{center}
\vspace{-0.5cm}
\end{table}

Ring is a high latency, bandwidth optimal algorithm, where each rank in the communicator sends data to the next rank and receives data from the previous rank. It offers maximum bandwidth by partitioning data into small chunks and pipelines them along the ring. For AllReduce, this setup leads to $2 \times (N - 1)$ sends and receives with size $S/N$, where $S$ is the size of the data to be reduced and $N$ is the number of ranks in the communicator. 

The tree algorithm was introduced in NCCL 2.4 to improve the scalability. It is a logarithmic latency algorithm which has a good performance on small and medium size operations \cite{SANDERS2009581}. It uses a double binary tree approach which pipelines a Reduce and a Broadcast to implement an AllReduce operation. Each rank in AllReduce primitive with tree algorithm sends and receives $2 \times S$ except the root, which is just $S$.     

The collnet algorithm allows GPUs on multiple nodes to do in-network reductions by using SHARP plugin \cite{sharp} for Mellanox switches. In-network reductions improve performance by eliminating the need to send data multiple times between endpoints.

\vspace{-0.4cm}

\section{Evaluation}
\label{sec:experiment}
We evaluate the results of our tool on two applications: a machine translation application, which uses Google's Neural Machine Translation model \cite{wu2016googles} and an image classification application, which employs a 18 layer Residual Neural Network (ResNet-18) model \cite{he2015deep}. A DGX-2 system with 16 NVIDIA Tesla V100 GPUs is used for evaluation. CUDA 10.1 and NCCL 2.7.8 are used for the experiments. The overhead of \tool{} for collective communication profiling is 1.4x on average. Since the prior work \cite{ComScribe} already shows the P2P capabilities of \tool{}, we mainly focus on collective communications in our evaluation. 

\subsection{Machine Translation Model}
\begin{figure}[t!]
\begin{center}
  \includegraphics[width=0.5\textwidth]{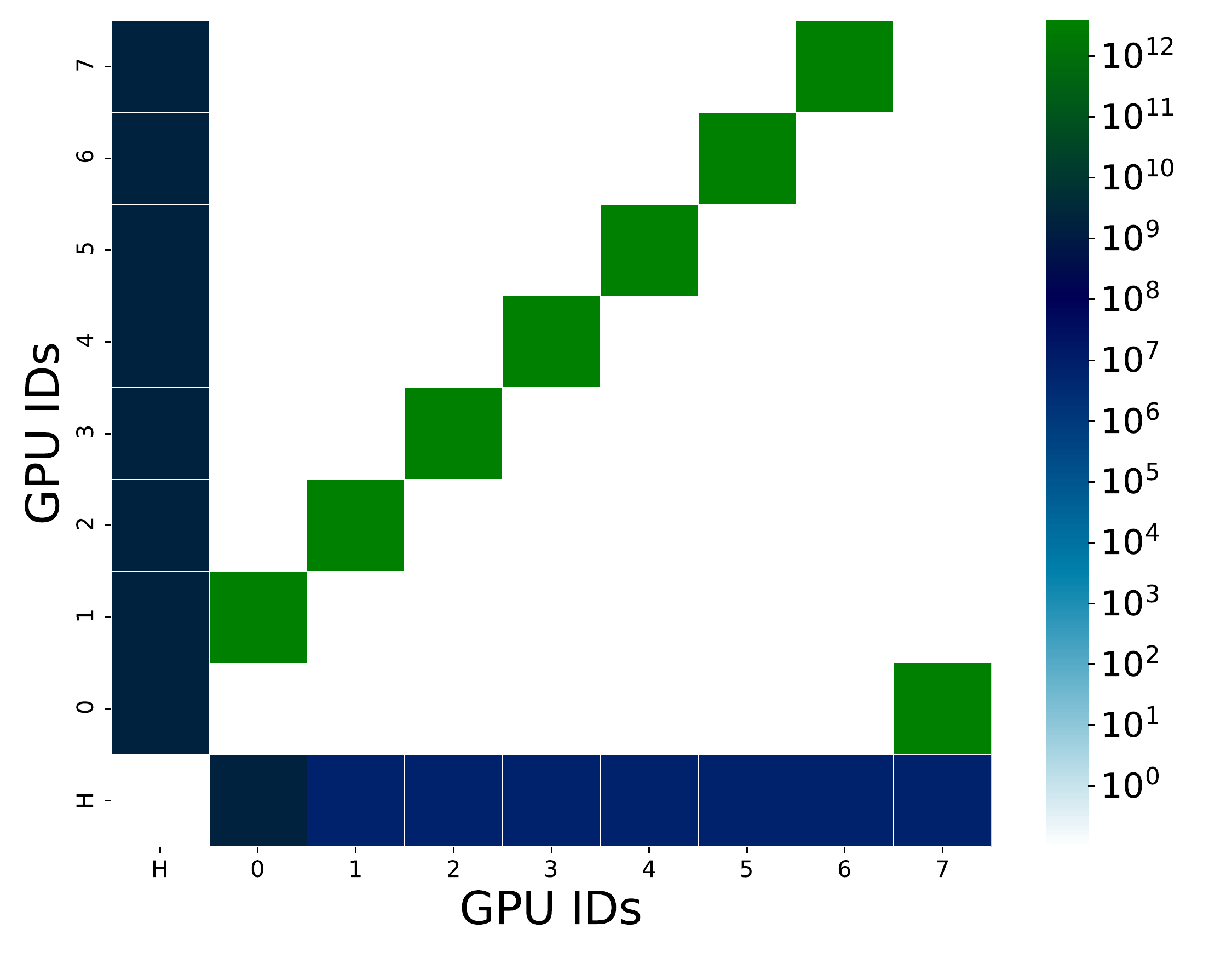}
  \end{center}
  \caption{Communication matrix of GNMT on 8 GPUs that shows the number of bytes transferred between CPU-GPU and GPU-GPU pairs for both P2P and collective communication. (0,0) is reserved for host.}
  \label{fig:gnmtall}
\vspace{-0.5cm}
\end{figure}

To demonstrate the capabilities of \tool{}, we profile a data parallel Google's Neural Machine Translation (GNMT) model with an improved attention mechanism \cite{nvidiadeeplearning} on WMT16 English-German dataset \cite{Koehn2005EuroparlAP}. 
Figure \ref{fig:gnmtall} shows the communication matrix of GNMT model for both P2P and collective communication combined in log scale. The communication matrix generated with \tool{} is a (\textit{d}+1)$*$(\textit{d}+1) matrix where \textit{d} is the number of GPUs. X- and Y-axis indicate the GPU ids. (0,0) entry is reserved for the host. Other entries in the matrix show the number of bytes transferred between a CPU-GPU or GPU-GPU pairs. 

Table \ref{tab:gnmttable} shows the number of calls made to each communication type and the amount of data movement  for each type. An interesting observation from the table is that the implementation of the GNMT model performs explicit transfers more than any other transfer types. Since explicit data transfer time is composed of a fixed latency and a component that is proportional to the transfer size, small sized transfers are dominated by the fixed latency. An optimization 
could be to bundle these fine-grained messages into more coarse-grained transfers. 




\begin{figure}[th]
  \begin{subfigure}{0.33\textwidth}
    \includegraphics[width=\linewidth]{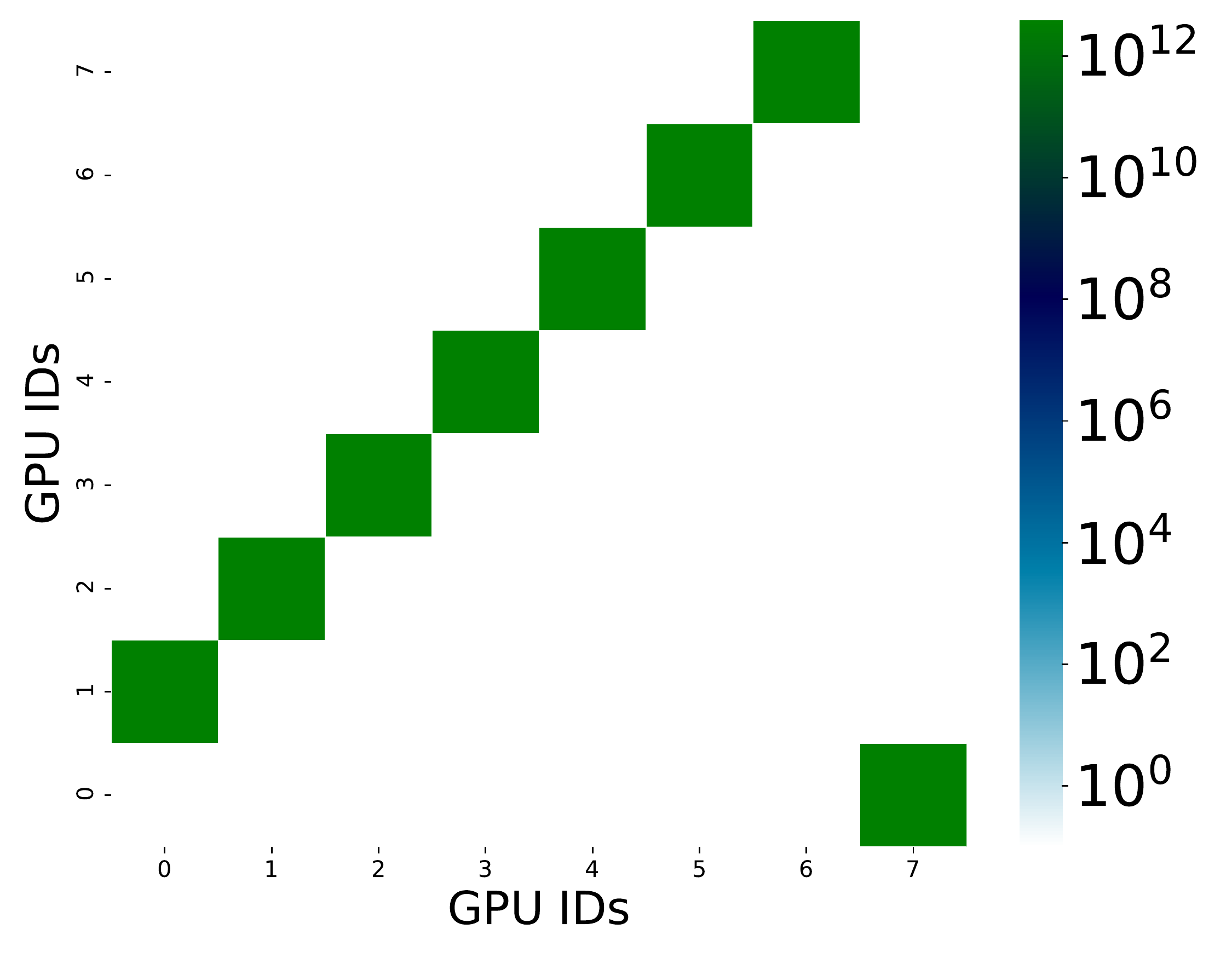}
    \caption{AllReduce} \label{fig:1a}
  \end{subfigure}%
  \begin{subfigure}{0.33\textwidth}
    \includegraphics[width=\linewidth]{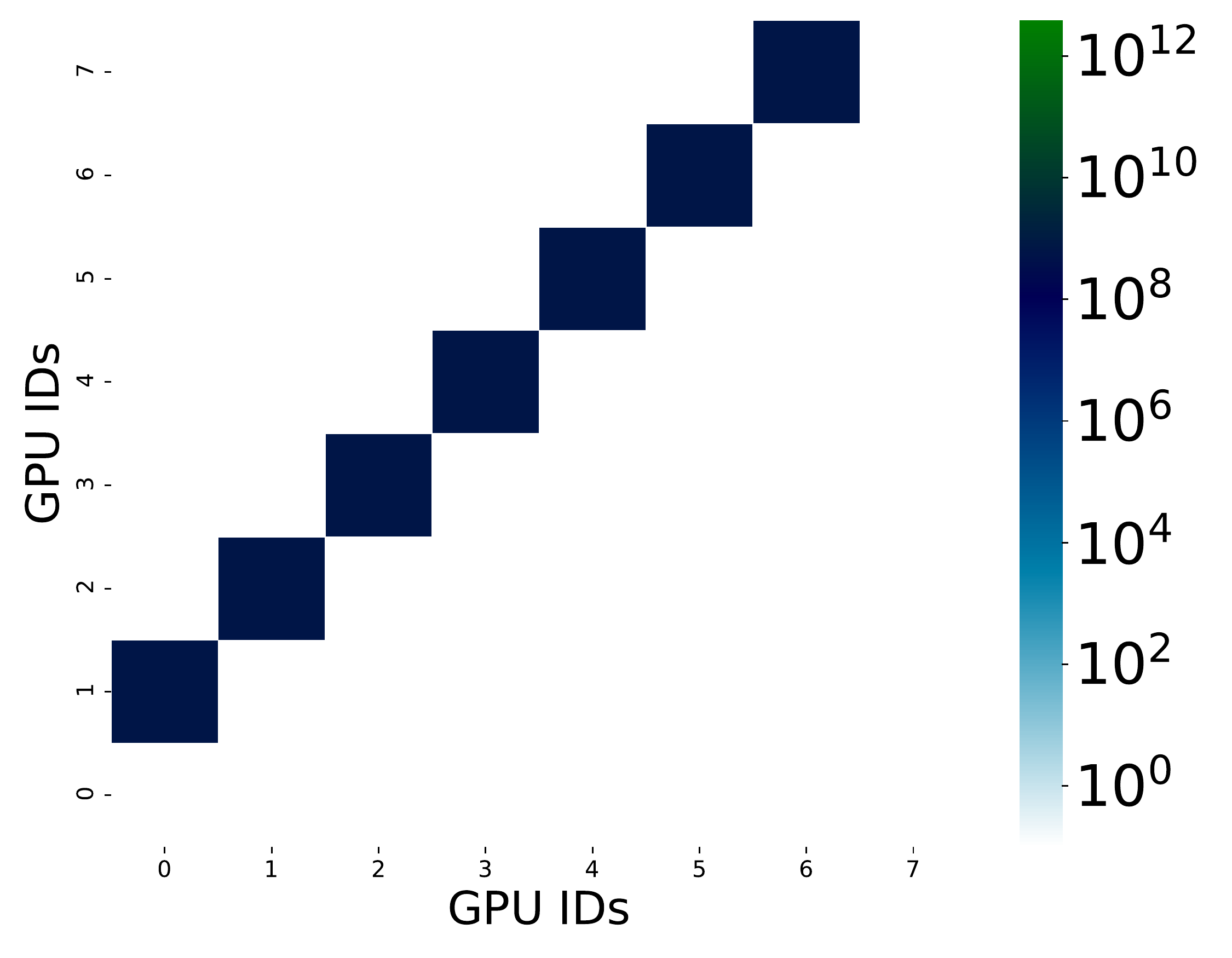}
    \caption{Broadcast} \label{fig:1b}
  \end{subfigure}
  \begin{subfigure}{0.33\textwidth}
    \includegraphics[width=\linewidth]{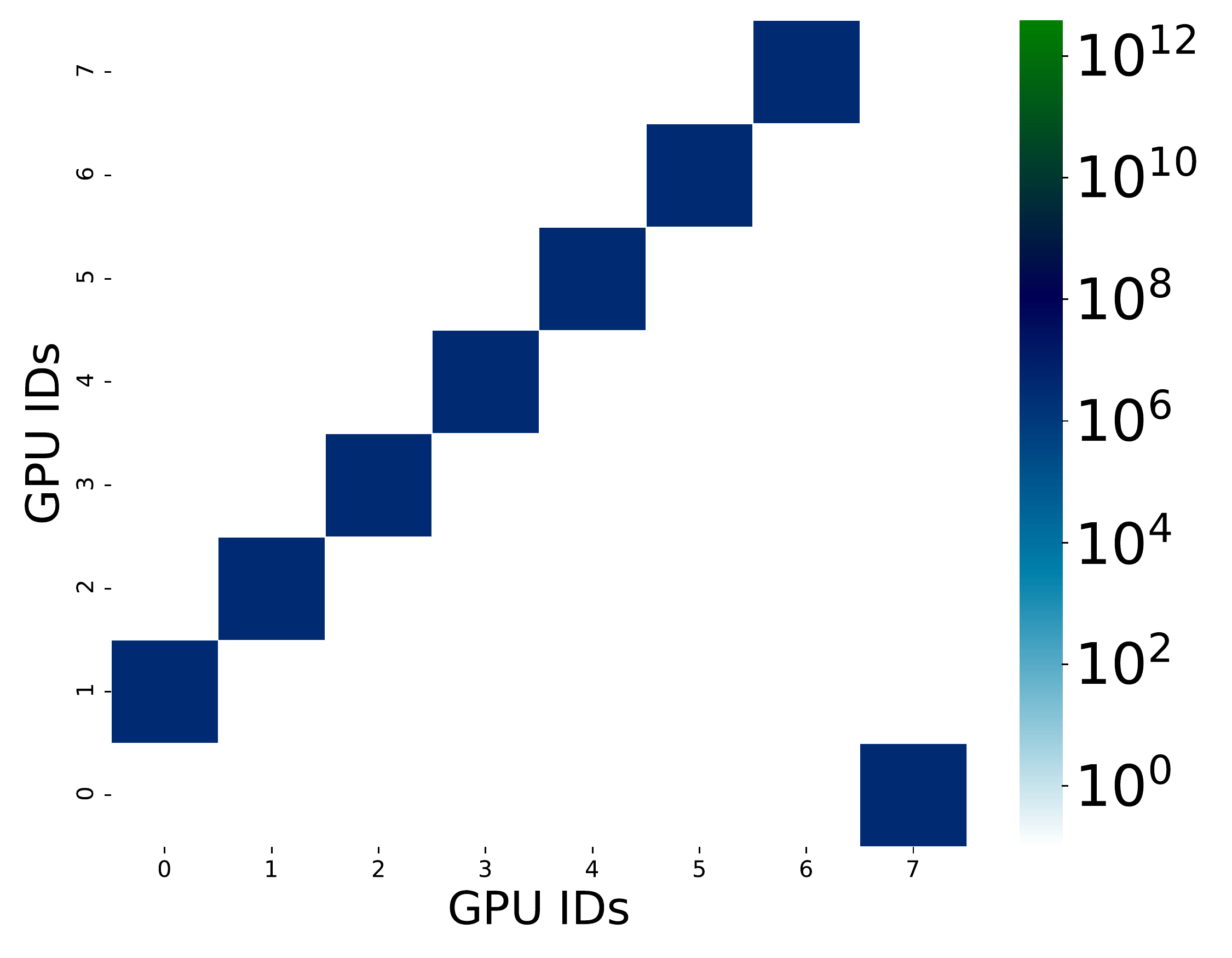}
    \caption{AllGather} \label{fig:1c}
  \end{subfigure}

\caption{Communication matrix for each collective that is used during the training of GNMT model. Number of bytes transferred with AllReduce on the left, Broadcast in the middle, and AllGather on the right in logarithmic scale}
\label{fig:GNMT-collectives}
\end{figure}

To better understand the usage of collective communication primitives, our tool can also produce matrices for each collective and P2P operation separately. The implementation of GNMT uses three collective primitives during the training of the machine translation model: AllReduce, Broadcast and AllGather. Figure \ref{fig:GNMT-collectives} shows that AllReduce operation is responsible for most of the collective communications. Hence, the time spent on optimizing AllReduce operation might have a good return on investment.

\begin{table}[h]
\scriptsize{
\caption{\label{tab:gnmttable}Communication primitive usage analysis of GNMT application. } 
\begin{center}
\begin{tabular}{ |c|c|c| } 
 \hline
 Communication  & Number of  & Total Size  \\ 
 Type & Calls & (in Mbytes) \\ 
 \hline
 AllReduce & $30739$ & $3,661,704$ \\
 \hline
 Broadcast & $5$ & $612$ \\
  \hline
 AllGather & $3$ & $3$ \\ 
 \hline
 Explicit Transfers & 778694 & $15,711$ \\ 
 \hline
 Unified Memory & 0 & 0 \\
 \hline
 Zero Copy Memory & 0 & 0 \\
 \hline
\end{tabular}
\end{center}
}
\vspace{-0.5cm}
\end{table}

\subsection{Image Classification Model}
Convolutional Neural Networks (CNNs) are widely used to classify images as they are capable of extracting various features from the given set of training images and infer the class of unseen images. We use a distributed data-parallel PyTorch implementation of ResNet-18 model with NCCL backend \cite{pytorch} to classify images on a subset of ImageNet \cite{5206848} dataset, which consists of 120000 images, where the size of each image is 64×64. 








In a data-parallel training, the data is first distributed across GPUs in the system and each GPU runs the same model on mini-batches from its own local data. Once each GPU completes its forward and backward passes independently in an iteration, a gradient exchange among GPUs occur to aggregate the gradients of the weights. However, there are various optimizations \cite{hermans2017accumulated,li2020pytorch,pytorchgradient} that can be implemented by library developers or the users of the libraries to complete the second step, which changes the collective communication frequency. 
For example, instead of launching AllReduce in every iteration to update gradients, the application can conduct a number of local training iterations before synchronizing gradients globally. Another optimization example that PyTorch implements is gradient bucketing \cite{li2020pytorch}, which increases throughput and decreases latency. Gradient bucketing method buckets multiple gradients into one \texttt{ncclAllReduce} call instead of sending each tensor separately.

\tool{} can help users to understand the effect of gradient bucketing on data movement. Table \ref{tab:imagenet} shows the number of calls to each NCCL primitive used during the training and the total size of the communication detected by \tool{}. A naive implementation of the gradient exchange step would be calling AllReduce operation for each parameter as soon as the gradient is ready for that parameter. In this naive approach, the number of AllReduce calls in an epoch would be equal to $DxN$, where $D$ is the number of parameters and $N$ is the number of iterations, yet since PyTorch implements gradient bucketing, the number of calls to the AllReduce operation is less than the naive approach. 

\setlength{\tabcolsep}{1em}
\begin{table}
\caption{\label{tab:imagenet}Number of execution of each primitive, total size used in ResNet-18 trained on a subset of ImageNet dataset for one epoch}
\begin{center}
\begin{tabular}{ |c|c|c| } 
 \hline
 Collective Operation & Number of Calls & Total Size (Bytes) \\ 
 \hline
 ncclAllReduce & 1174 & $3.2 \times 10^{10}$ \\
 \hline
 ncclBroadcast & 789 & $6.1 \times 10^7$ \\
  \hline
\end{tabular}
\end{center}
\end{table}

\vspace{-0.5cm}

\vspace{-0.2cm}
\section{Related Work}
There are several tools that can trace memory transfers of host-device and device-device pairs with LD\_PRELOAD utility (EZTrace \cite{eztrace}, Extrae \cite{extrae}, and Score-P \cite{scorep}). These tools can generate execution traces for various programming models including MPI, OpenMP, CUDA, and PThread. However, the profiling support for CUDA memory transfer functions is limited with explicit memory transfer types (i.e. \textit{cudaMalloc} and \textit{cudaMemcpy}) and NCCL tracing is not supported by any of them. Our tool can detect collective communication primitives of NCCL and various P2P communication types such as Unified Memory and Zero-Copy memory.   


Tartan, multi-GPU benchmark suite \cite{tartan1,tartan2}, consists of micro-benchmarks and applications to evaluate the performance of modern interconnects such as PCIe, NVLink 1.0, NVLink 2.0, NV-SLI, NVSwitch and Infiniband systems with GPUDirect RDMA in scale-up (intra-node) and scale-out (inter-node) scenarios.  
Even though Tartan assesses interconnect performance in terms of latency, bandwidth, and efficiency on message size for P2P and collective communications, 
it is not a tool that can be used to monitor and detect communications of an application.

Nsight Systems is NVIDIA's visualization tool that aims to help users to identify potential optimizations for their applications. It can provide a timeline of the executed functions and data transfer information for CUDA memory operations. With 2020.5 and 2021.2 releases, NCCL support was added for timeline visualization but currently it does not show the underlying communication among GPUs. 
Our tool can log communication among GPUs for collective NCCL calls in a machine readable format whereas to our knowledge Nsight Systems command line interface can only show the time it takes to run a single collective call at the moment.

Scope \cite{commscope} is a benchmark framework which consists of various benchmark suites such as Comm\textbar Scope, NCCL\textbar Scope and many others. Comm\textbar Scope is a NUMA-Aware multi-CPU multi-GPU benchmark suite that measures point-to-point transfer latency and bandwidth within a single node for different data transfer scenarios with CUDA P2P communication types such as Unified Memory and Zero-copy Memory. NCCL\textbar Scope consists of micro-benchmarks to measure the bandwidth of all five NCCL primitives with \textit{cudaEvent}. Even though our work and Scope have features in common such as the categorization of communication types, our work supports the recording of communication for any application. 

There are number of tools to generate communication patterns for multi-core applications. ComDetective \cite{comdetective} detects inter-thread data transfers by using debug registers and Performance Monitoring Units for multi-threaded applications. 
Similar to ComDetective, Azimi et al. \cite{azimi2009enhancing} and Tam et al. \cite{tam2007thread} use kernel support to access PMUs and the kernel generates the communication pattern for the applications. Simulator-based approaches to collect memory access traces for generating communication patterns include Barrow-Williams et al. \cite{barrow2009communication} and Cruz et al. \cite{da2011using}. Numalize \cite{diener2015characterizing,diener2016communication} uses binary instrumentation to intercept memory accesses and captures communication between threads accessing the same address in memory. None of the aforementioned tools, however, have support for multi-GPU communication. 

\label{sec:related}

\vspace{-0.3cm}
\vspace{-0.2cm}
\section{Conclusion}
The communication among GPUs is a critical performance and scalability contributor in multi-GPU systems. \tool{}, our prior work, identifies and analyzes implicit and explicit P2P communication types. This work extends \tool{} to support collective communication profiling for GPUs sharing a common host. 
To implement the collective communication support in \tool{} we take advantage of LD\_PRELOAD utility to identify and extract the communication among GPUs in a communicator. 
We evaluated our tool against two deep learning applications. Our tool can provide insights to study the communication patterns of collective operations.

\vspace{-0.3cm}
\section*{Acknowledgement}
The work is supported by the Scientific and Technological Research Council of Turkey (TUBITAK), Grant no. 120E492. Dr. Didem Unat is supported by the Royal Society-Newton Advanced Fellowship.
\vspace{-0.3cm}

{\scriptsize{
\bibliographystyle{splncs04}
\bibliography{references}
}}
\end{document}